\documentstyle[11pt,paspconf,epsf]{article}
\def\ref{\par\noindent\hangindent=1truecm}
\font\piedi=cmr8

\catcode`\@=11
\def\gsim{\ifmmode{\mathrel{\mathpalette\@versim>}}
    \else{$\mathrel{\mathpalette\@versim>}$}\fi}
\def\lsim{\ifmmode{\mathrel{\mathpalette\@versim<}}
    \else{$\mathrel{\mathpalette\@versim<}$}\fi}
\def\@versim#1#2{\lower 2.9truept \vbox{\baselineskip 0pt \lineskip 
    0.5truept \ialign{$\m@th#1\hfil##\hfil$\crcr#2\crcr\sim\crcr}}}
\catcode`\@=12

\def\yr-1{\hbox{${\rm yr}^{-1}$}}

\def\mg2{\hbox{${\rm Mg}_2$}}
\def\msun{\hbox{$M_\odot$}}

\def\ho{\hbox{$H_\circ$}}
\def\h50{\hbox{$\ho /50$}}
\begin{document}

\title{The Star Formation History of Ellipticals from the Fossil Evidence}
\author{Alvio Renzini}
\affil{European Southern Observatory, D-85748 Garching b. M\"unchen, Germany}
\author{Andrea Cimatti}
\affil{Osservatorio Astrofisico di Arcetri, Largo E. Fermi 5, I-50125, Firenze,
       Italy}




\begin{abstract}
The current evidence about the age of stellar populations in elliptical 
galaxies is reviewed. The case for the bulk of stars in galactic spheroids
(ellipticals and bulges) having formed at high redshift ($z\gsim 3$)
is now compelling, both for cluster and well as field galaxies. Whether 
the assembly of ellipticals is deferred to lower redshifts compared to the
formation of their stars remains controversial, and we mention ongoing 
observational programs that are designed to solve the issue.
We present preliminary results
of a pilot project aimed at ascertain the nature of extremely red galaxies
(ERGs) with near-IR spectroscopy at the VLT. 

\keywords{elliptical galaxies, clusters of galaxies, high redshift galaxies,
galaxy formation}
            
\end{abstract}      



\section{The Age of Cluster Ellipticals}
Through the '80s it became normal to start a talk about ellipticals by saying
that `the classical view of elliptical (and spheroid) formation was that they 
formed stars efficiently in the early universe, and then their 
stellar population evolved passively after this initial burst'. Then 
the  talk would have continued by saying `... however, in recent years
there has been growing evidence that this is not the true story, ellipticals
are rather {\it intermediate age} objects, and star formation in them 
has continued for a major fraction of the Hubble time and in some cases 
is still going on today ...'. Actually, such evidence hardly grew, in spite of
a widespread expectation given the success of
hierarchical theories of structure formation coupled to the aesthetic appeal 
of $\Omega=1$ inflationary cosmology.

>From the turn of the decade through all the '90s the has been instead
growing empirical 
evidence that ellipticals are dominated by very old stellar
populations. The first breakthrough came from noting the tightness of the
color-$\sigma$ relation of ellipticals in the Virgo and Coma
clusters (Bower, Lucey, \& Ellis 1992). This demands a high degree of
{\it synchronization} in the star formation history of these ellipticals, that
is most naturally accounted for by pushing back to early times most of
the star formation. Making minimal use of stellar 
population models,  this approach provided for the first time
a {\it robust} demonstration that at least
{\it cluster} ellipticals are made of very old stars, with the bulk
of them having formed at $z\gsim 2$. 

The main lines of the Bower et al. argument are as follows. 
The observed color scatter of cluster ellipticals is related to
the age dispersion among them by the relation:
\begin{equation}
\delta (U-V) = {\partial (U-V)\over\partial t}(t_{\rm H}-t_{\rm
F})
\end{equation}
where $t_{\rm H}$ and $t_{\rm F}$ are  the age of the ``oldest''
and ``youngest'' galaxies, respectively. Here by age one intends the 
luminosity-weighted age of the stellar populations that constitute such
galaxies. The time derivative of the color is obtained from evolutionary
population synthesis models, which give $\partial (U-V)/\partial
t\simeq 0.02$ mag/Gyr for $t\simeq 10$. The observed color scatter is
$\delta (U-V)\simeq 0.04$ mag, consistent with pure
observational errors. Hence, one gets $t_{\rm H}-t_{\rm F}\lsim 0.04/0.02=2$
Gyr, and if the oldest galaxies are 15 Gyr old, the youngest ones
ought to be older than 13 Gyr, from which Bower et al. conclude they
 had to form at $z\gsim 2$. If the oldest galaxies were instead  as young as, 
say 5
Gyr, then the youngest should be older than at least 3 Gyr, which
would require a high degree of synchronization in their formation, which
seems unlikely.

Evidence in support of the Bower et al.
conclusion has greatly expanded, to finally become compelling. It is worth
mentioning some of the steps of this development:
the tightness of the fundamental plane
relation for ellipticals in local clusters (Renzini \& Ciotti 1993),
the tightness of the color-magnitude relation for ellipticals in
clusters up to $z\sim 1$ (e.g., Aragon-Salamanca et al. 1993; Taylor
et al. 1998; Kodama et al. 1998; Stanford,
Eisenhardt, \& Dickinson 1998), and the
modest shift with increasing redshift in the zero-point of the fundamental
plane, Mg$-\sigma$, and color-magnitude relations of cluster
ellipticals (e.g., Bender et al. 1997;
Dickinson 1995, 1997; Ellis et al. 1997; van Dokkum et al. 1998;
Pahre, Djorgovski, \& de Carvalho 1997; Stanford, et al.
1998; Kodama et al. 1998). All these studies
agree in concluding that most stars in ellipticals formed at $z\gsim
3$, though the precise lower limit of the redshift
 depends on the adopted cosmology; it would be more  like $z\gsim 2$ in the 
current {\it standard} cosmology: $(\Omega,\Lambda)=(0.3,0.7)$.

It is worth emphasizing that all these studies have in common the same 
methodology pioneered by Bower et al. (1992). They focus indeed on
the tightness of  some
correlation among the global properties of cluster ellipticals, which
sets a robust constraint on their age dispersion as
opposed to an attempt to
date individual galaxies. Moreover, the move to high redshift offers two
fundamental advantages. The first advantage is that looking at high
$z$ provides the best
possible way (we should say {\it the} way) of removing the
age-metallicity degeneracy. If spheroids are made of intermediate-age,
metal rich stars, they should become rapidly bluer and then disappear
already at moderate redshift, which is not the case
 (e.g. Kodama \& Arimoto 1997). 

It has also been demonstrated that the scatter about the average
Mg$-\sigma$ relation (among local $z\simeq 0$ cluster ellipticals) is larger
than expected from pure observational errors (Colless et
al. 1999). Colless et al. interpret this as evidence for a substantial
age dispersion, but it remain to be seen whether this interpretation is
consistent with the color scatter remaining constant all the way to 
$z\simeq 1$, as found by Stanford et al. (1998).
The observational opportunity of studying galaxies at
large lookback times makes quite obsolete those attempts to find 
combinations of spectral indices that are aimed at distinguishing
between age and metallicity effects in nearby galaxies. The obvious drawback
of these indices (e.g. H$_\beta$) is that they are very sensitive to
even minor recent episodes of star formation, and therefore they do not
help to determine the formation epoch for the {\it bulk} of
stars in ellipticals.

The second advantage of high$-z$ studies is that at high redshift one
gains more {\it leverage}: for given dispersion in some observable one
can set tighter and tighter limits to the age dispersion. This comes
from the color time derivatives being larger the younger the
population. For example, the derivative $\partial (U-V)/\partial t$ is
$\sim 7$ times larger at $t=2.5$ Gyr than it is at $t=12.5$ Gyr
(e.g. Maraston 1998), and therefore a given dispersion in this
rest-frame color translates into a $\sim 7$ times tighter constraint
on age and therefore on formation redshift.  The case is effectively
illustrated by isolated high redshift ellipticals: Spinrad et
al. (1997) found a {\it fossil} (i.e. passively evolving) elliptical
at $z=1.55$ for which they infer an age of at least 3.5 Gyr, hence a
formation redshift in excess of $\sim 5$. A similar case is
anticipated by Dunlop (1998), with a galaxy at $z=1.43$ and an age of
3-4 Gyr.  At an even higher formation redshift may lie the extremely
red galaxy in the NICMOS field of the HDF-South, which spectral energy
distribution is best accounted for by an old, passively evolving
population at $z\simeq 2$ (Stiavelli et al. 1999).

\section{Cluster vs Field Ellipticals}

Most of the evidence mentioned in the previous Section refers
to cluster ellipticals.  In hierarchical models, clusters form out of
the highest peaks in the primordial density fluctuations, and cluster
ellipticals completing most of their star formation at high redshifts
could be accommodated in the model (e.g. Kauffmann 1996; Kauffmann \&
Charlot 1998a). However, these models predict that in lower density, 
{\it field} environments,
both star formation and merging should be appreciably delayed to later times,
which offers the opportunity for an observational
test of the hierarchical merger paradigm.

The notion of field ellipticals being a less homogeneous family compared
to their cluster counterparts has been widely entertained, though -- once
more - the
direct evidence has been only rarely discussed.  Visvanathan \& Sandage
(1977) found cluster and field ellipticals to follow the same
color-magnitude relation, but Larson, Tinsley, \& Caldwell (1980) --
using the same database -- concluded that the scatter about the mean
relation is larger in the field than in clusters.
More recently, a larger scatter in field versus
cluster ellipticals was also found for the fundamental plane 
relations by de Carvalho \& Djorgovski (1992). However, when using absolute
luminosities and effective radii
at least part of the larger scatter among field ellipticals certainly comes
from their distances being more uncertain than for cluster galaxies.

Taking advantage of a large sample ($\sim 1000$) of early-type
galaxies with homogeneously determined Mg$_2$ index and central
velocity dispersion, Bernardi et al. (1998) have recently compared the
Mg$_2-\sigma$ relations (which are distance independent!) of cluster
and field galaxies. The result is that field and cluster ellipticals
all follow basically the same Mg$_2-\sigma$ relation. The zero-point
offset between cluster and field galaxies is $0.007\pm 0.002$ mag,
with field galaxies having lower values of \mg2, a statistically
significant, yet very small difference.  This is in excellent
agreement with the offset of $0.009\pm 0.002$ mag, obtained by
J\/orgensen (1997) using 100 field and 143 cluster galaxies.

Using the time derivative of the Mg$_2$ index from synthetic stellar 
populations, Bernardi et al. conclude that the age difference between
the
stellar populations of cluster and field early-type galaxies is at
most $\sim 1$ Gyr. The actual difference in the mass-weighted age
(as opposed to the luminosity-weighted age) could be significantly
smaller that this. It suffices  that a few galaxies have undergone a minor star
formation event some Gyr ago, with this having taken place
preferentially among field galaxies.

\section{Spiral Galaxy Bulges}

According to general wisdom, the bulges of spiral galaxies are hard to 
distinguish from elliptical galaxies of the same luminosity, once they are
stripped of the disk which rotates around them. This is illustrated for 
example by most bulges following the same Mg$_2-\sigma$ and Mg$_2-$luminosity
relation as common ellipticals (Jablonka, Martin, \& Arimoto 1996).
In their sample just $\sim 20\%$ of the studied bulges appear to be
appreciably bluer that the standard relation for ellipticals, a sign that 
recent star formation is required in only a minority of the bulges.

Closer to us, contrary to several early claims, no evidence exists for an 
intermediate age population for the bulge of M31 
(e.g. Renzini 1998; 1999; Jablonka et al. 1999). Even more stringently,
HST and ground based color-magnitude diagrams of Galactic bulge globular 
clusters and fields indicate a uniform old age for our own bulge
(Ortolani et al. 1995).
Given that we leave in a galaxy that is member of a rather loose group,
certainly not an exceptionally high peak in the primordial density 
fluctuations, it seems reasonable to conclude that star formation in most
galactic spheroids, ellipticals and bulges alike, was essentially complete
at high redshift, no matter whether such spheroids resides today in high- or 
low-density environments.

\section{Do Ellipticals Disappear by $z\simeq 1$?}
In apparent conflict with this conclusion are several claims according to which
the comoving number density of field elliptical galaxies is rapidly 
 decreasing 
with redshift by $z\simeq 1$ (e.g. Kauffmann, Charlot, \& White 1996; Zepf 
1997; Meneanteau et al. 1998; Franceschini et al. 1998). 
This is a controversial result. Other groups claim such dismissal of 
high-redshift ellipticals to be premature, either because the adopted models
for putative high-$z$ ellipticals would be too naive (Jimenez et al. 1999),
or because they
find evidence for the comoving number density of ellipticals 
being constant to at least
$z\sim 1$ (e.g. Totani \& Yoshii 1998; Benitez et al. 1998; Shade et al. 
1999), and possibly as much as to $z\sim 2$ (Broadhurst \& Bouwens 1999).

Part of this discrepancy is likely to arise from the samples so far analyzed 
being rather small, both in terms of the total number of galaxies involved, 
and especially in terms of the (small) number of large scale structures that
happen to be included in each sample (see below). Part of the discrepancy
may also arise from the use of different selection criteria for ellipticals.
Clearly, these differences need to be understood before claiming evidence in 
support of a late assembly of ellipticals (as favored by most current 
realization of the CDM model of hierarchical formation of cosmic structures,
e.g. Kauffmann \& Charlot 1998a), or in support of an early build-up of massive
galaxies, resembling the ``old-fashioned'' monolithic collapse models.

\section{Searching for Fossil Ellipticals at High $z$: 
Early VLT Observations of Extremely Red Galaxies}

The existence and the abundance of fossil ellipticals at $z>1$ 
can be constrained with the selection and study of faint galaxies 
with the colors expected for passively evolving old stellar populations. 
For instance, a color threshold of $R-K \gsim 5.3$ is expected at 
$z \gsim 1$ in the case of a passively evolving, solar metallicity
stellar population with $z_{\rm form}>3$, $SFR \propto 
exp(-t/\tau)$ and $\tau$=0.1 Gyr (adopting the Bruzual \& Charlot 1998 
models, and assuming $H_0=50$ kms$^{-1}$ Mpc$^{-1}$ and $\Omega_0=1$).
The observed colors of the few known isolated $z>1$ ellipticals (see 
Section 1) are indeed consistent with those expected from the spectral 
synthesis models. 

Do such  color-candidate ellipticals at $z>1$
exist? Observationally, the answer is yes: a population of extremely 
red galaxies (hereafter ERGs) was discovered with the combination of 
optical and near-IR imaging (e.g. Elston et al. 1988; McCarthy et al. 
1992; Hu \& Ridgway 1994). ERGs are ubiquitously found in empty sky 
fields (Cohen et al. 1999; Thompson et al. 1999), in the vicinity of 
high-$z$ AGN (McCarthy et al. 1992; Hu \& Ridgway 1994) and as counterparts 
of faint X-ray (Newsam et al. 1997) and weak radio sources (Spinrad et al. 
1997). Their typical magnitudes $K\sim 18-20$, and their surface density 
for $R-K>6$ and $K<19$ is 0.039$\pm0.016$ ERGs/sq. arcmin (Thompson et 
al. 1999). Their nature is still poorly known, as their faintness makes 
optical and near-IR spectroscopic observations quite difficult,
but they are likely to represent a {\it mixed bag} class of objects.
Indeed, their red colors are consistent with them being passively evolving 
ellipticals,  but also with strong dust 
reddening in a star-forming or active galaxy. A dramatic example of 
this ambiguity is provided by HR10: its spectral energy distribution 
(SED) is consistent with that of an old elliptical at $z\approx2.4$ 
(Hu \& Ridgway 1994), but near-IR and submm observations proved that 
HR10 is a dusty starburst galaxy at $z=1.44$ (Graham \& Dey 1996; 
Cimatti et al. 1998; Dey et al. 1999). On the other hand, recent 
observations suggest that the ERG CL 0939+4713B ($R-K\sim7$) is 
an old galaxy at $z\sim1.6$, based on the detection of the redshifted 
4000~\AA~ break (Soifer et al. 1999). 

In order to investigate the nature of ERGs and their role in the
framework of elliptical galaxy formation and evolution, we recently
started a project based on optical and near-IR imaging and spectroscopy
with the ESO NTT 3.5m and the VLT 8m telescopes. 
Near-infrared ($JHK$) spectroscopy of an incomplete ``pilot'' sample 
of ERGs was made in April 1999 with the ESO VLT-UT1 ({\it Antu}) equipped 
with the IR imager-spectrograph ISAAC (Moorwood et al. 1999). The typical 
integration times were 1-2 hours per spectrum and the resolution was 
$\sim 500$. Nine ERGs with $R-K>5$ 
or $I-K>4.5$ were observed with the main purpose of finding cases of 
{\it bona fide} ellipticals at $z>1$ or, in case of dusty starburst/active 
galaxies, to search for redshifted emission lines (such as H$\alpha$ at 
$z>0.7$) (Cimatti et al. 1999). 

\begin{figure}
\plotfiddle{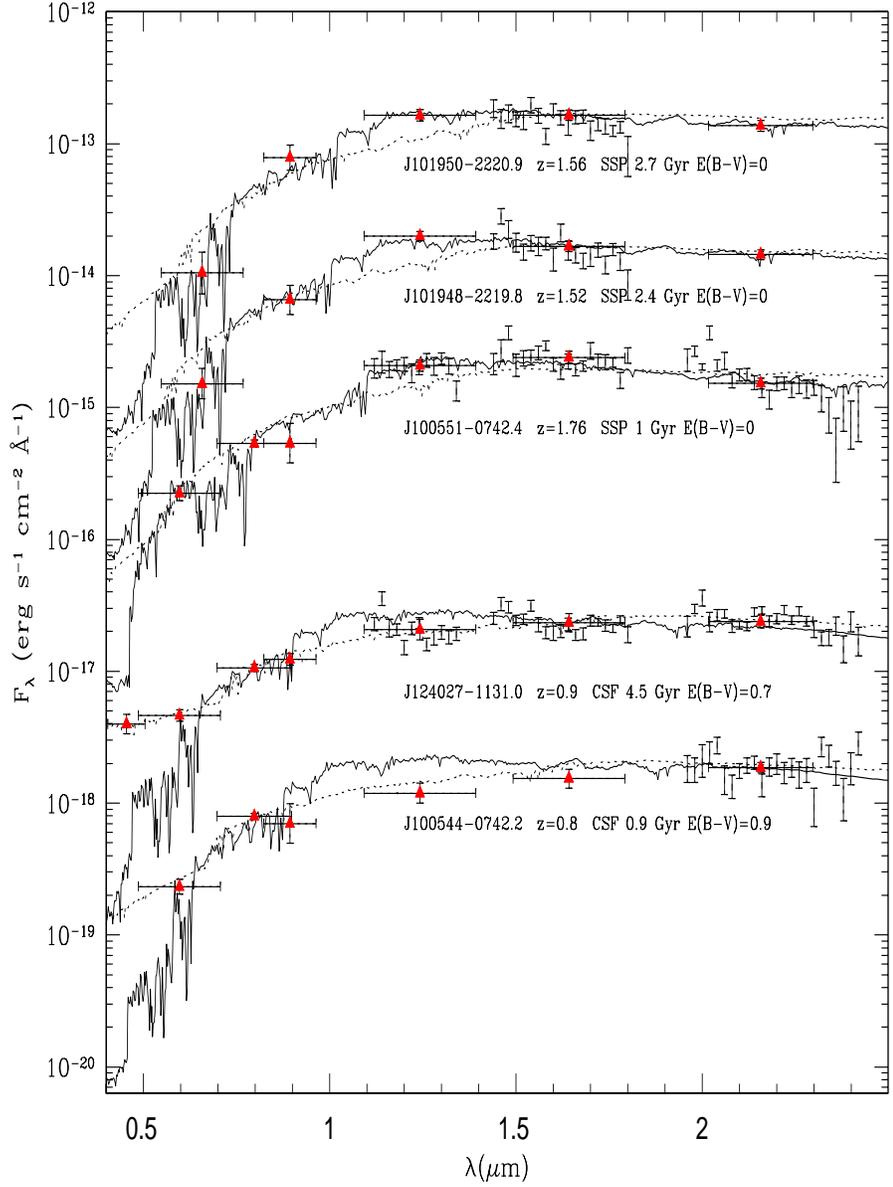}{9truecm}{0}{65}{90}{-180}{-290}
\vskip4.9truecm
\caption{\piedi The optical to near-IR SEDs of five ERGs from broad-band 
photometry (filled triangles) and ISAAC near-IR spectroscopy (vertical 
bars). The spectra are binned with 200~\AA~ wide bins in order to reach 
a typical $S/N\approx$10 per bin. The first three SEDs from the top are 
those consistent with both passively evolving ellipticals and no dust 
extinction. Instead, the other two SEDs require substantial dust
extinction to be reproduced. The continuum and the dotted lines are 
the best fit spectra  obtained respectively with dustless, old stellar 
populations and with dust-obscured constant star formation. The best 
fit parameters ($z$, ages and $E_{B-V}$) are shown relatively to the 
best fit of the two cases (i.e. dusty or dustless). From bottom to top, 
the fluxes of the five SEDs are multiplied by $10^{0,1,2,4,5}$ respectively. 
}
\vfill
\end{figure}

 Continuum emission was detected in all the targets, but neither strong 
emission lines nor evident continuum breaks were detected in the ISAAC 
spectra. The depth of our spectra (typically $F_{\rm lim}<$
2$\times10^{-16}$ erg s$^{-1}$cm$^{-2}$) would allow to 
detect emission lines such as H$\alpha$ if as strong as in HR10 (Dey et al. 
1999). These limits on the H$\alpha$ emission  imply upper limits on
the star formation rates of individual galaxies that -- in absence
of dust extinction -- range from
$\sim 7$ to $\sim 30$$h_{50}^{-2} $M$_{\odot}$yr$^{-1}$ ($\Omega_0=1.0$). 
These limits  correspond to the SFRs as high as those of nearby 
gas rich spiral galaxies (Kennicutt 1998). However, 
in case of strong dust extinction, the SFRs would
increase significantly. For instance, adopting the Calzetti
(1997) extinction curve, for $E_{B-V}=$0.2--0.9, the corresponding
correction factors of the $SFR$ (as derived from H$\alpha$) are 
$\approx$2--30$\times$. 

In absence of distinctive spectral features, the ISAAC spectrophotometry 
combined with the available optical and near-IR broad-band photometry 
was used to estimate the ``spectrophotometric'' redshifts of the observed 
ERGs and to study their SEDs. We found that a fraction of the program
ERGs have SEDs 
consistent with being passively evolving ellipticals at $z\sim1.5-1.8$ 
with no dust extinction and rather old ages ($\sim 1-3$ Gyr) (see Fig.
1). Their rest-frame $K$-band absolute magnitudes ($M_{K}\sim$-24.8$\div$
-25.2) imply luminosities $L \lsim L^{\ast}$ (adopting $M_{K}^{\ast}$=
-25.16 for the local luminosity function of elliptical galaxies; Marzke et 
al. 1998). In addition, deep HST optical
imaging shows that their morphologies are compact and regular, consistent 
with such ERGs being high-$z$ ellipticals. 
However, in 
some cases the SEDs seem to require strong dust extinction 
to be reproduced (see Fig. 1), and in a fraction of these cases the 
morphologies are clearly 
disturbed (i.e. incompatible with being ellipticals). In other cases
the required dust may be an artifact of the fitting procedure, in 
which the metallicity of the population is kept fixed to solar.

On the one hand, our results show the existence of a population of faint
galaxies that are selectable through their very red colors and which 
have properties consistent with the strict definition of being dustless, 
old and passively evolved spheroidals at $z>1$ (see also Spinrad et al. 
1997; Soifer et al. 
1999). Not surprizingly, our observations also indicate that a non-negligible 
fraction of the ERG population consists of galaxies whose SEDs and 
morphologies are more consistent with being high-$z$, star-forming
 dusty systems.

The sample for this pilot study is clearly too small to draw more
quantitative conclusions at this stage. We now plan
somewhat deeper spectroscopic observations of a complete sample, with
the aim  to assess what 
fraction of the ERG population consists of passively evolving high-$z$ 
ellipticals and thus to compare their observed abundance with that
expected from pure luminosity evolution models.

\section{Do Bright Galaxies Disappear by $z\sim 1$?}

Generally, ellipticals have been selected either according to color or
morphological selection criteria, or to some combination
thereof. However, as one moves to high redshift, minor residual star
formation may cause virtually fully assembled ellipticals to drop out
of samples that are selected using color and/or morphological
information. To circumvent this limitation, Broadhurst, Ellis, \&
Glazebrook (1992) proposed to select objects only according to their
$K$-band magnitude and to construct their redshift distribution. This
provides a more robust measure of the evolution (if any) of the
comoving number density of massive galaxies, and a more fundamental
check of the merging paradigm. The
advantage of a $K$-band selection is that, as opposed to e.g.
$B$-band selected samples,  it favors massive
galaxies even if currently experiencing very low levels of star
formation.

Following this approach, Kauffmann \& Charlot (1998b) estimate that in their
 pure luminosity evolution (PLE) model $\sim 60\%$ of the galaxies in a
$K\le 20$ sample should be at $z>1$, while only $\sim 10\%$ are expected in
a standard hierarchical merging model. The largest redshift surveys of $K<20$
 selected samples currently available are the Keck samples of Cowie et al. 
(1996) covering $\sim 26$ square arcmin, and 
Cohen et al. (1998) covering $\sim 14$ square arcmin. Both spectroscopic
samples are rather incomplete, for example, the Cohen et al.
sample includes 195
objects, among which 24 are stars, 21 are galaxies at $z>1$, and optical
spectroscopy failed to provide a redshift for another 32 objects. The vast
majority of these latter objects are likely to be galaxies at $z\gsim 1$, 
in which
major spectral features have moved to the near IR. All in all, in this sample 
 $\sim 53/171\simeq 30\%$ of galaxies are likely to be at $z>1$.
This empirical value falls (ironically enough) just midway between
the prediction of the two competing models. Most importantly, Cohen et al.
emphasize that $\sim 50\%$ of their galaxies fall in just five `redshift peaks'
most likely due to clustering. Therefore, a much larger sample is required to
dispose of a `fair' representation of the galaxy population, one that is not 
subject to statistical fluctuations in the number of included large scale
structures. 

The conclusion is that the two Keck redshift survey of $K$-limited samples
are so far insufficient to conclude the experiment envisaged by Broadhurst, 
Ellis, \& Glazebrook (1992). Big samples are required, distributed over 
several different lines of sight so as to average out the effects of 
clustering and large scale structures. A contribution in this direction will
come from an observing program with FORS1, FORS2 and ISAAC at the VLT
(PI A. Cimatti), aimed at gathering redshifts of two $K<20$ samples over two
separate lines of sight, for a total of $\sim 500$ galaxies.

\section{Speculations}

In the previous sections it has been documented that compelling evidence 
exists for the bulk of stars in galactic spheroids being
very old, i.e. formed at redshifts beyond $\sim 3$, and possibly even
much beyond this value. This applies to 
ellipticals and bulges alike, in clusters as well as in the lower
density regions still inhabited by spheroids, including the bulge of
our own Galaxy. This is what was expected
(actually postulated) in the monolithic collapse scenario, while it 
is at variance with most existing realizations of the hierarchical merging
 scenario.

The fact that spheroids are made of old stars does not necessarily 
invalidate the hierarchical merging
paradigm, which actually offers a still unique description on how
large galaxies could have been assembled. In an effort to comply 
with the observations, 
hierarchical models should be tuned to mimic the monolithic model as much
as possible, primarily
by pushing most of the action back to an earlier cosmological epoch. 
This is certainly favored by low-$\Omega$ + $\Lambda$ cosmologies, but will
not suffice.
As universally recognized, the pitfall of semianalytical realizations of the
hierarchical paradigm is represented by how star formation and its feedback
are parameterized in the models. The canonical assumption has been that most 
of the star formation takes place in spiral disks, and ellipticals are formed
by merging spirals. We think that this widely entertained paradigm is far from
being proven. All we know about ellipticals  demands 
``that star formation, 
metal enrichment, merging and violent relaxation did not take place
sequentially; they more likely were all concomitant processes, along with
supernova heating and radiative cooling in a multiphase ISM'' (Renzini 1994).

With most of the merging taking place at high
redshifts, among still mostly gaseous components, 
merging itself would promote the widespread starburst activity responsible
for the prompt buildup of galactic spheroids (Somerville \& Primack 1998;
Renzini 1999). The natural 
observational counterparts of these events may be represented by the
Lyman-break galaxies at $z\gsim 3$ (Steidel et al. 1999), where star
formation rates could be in extreme cases as high as $\sim 1000\; \msun\yr-1$ 
(Dickinson 1998). It remains to be explored whether such tuning of the
star formation algorithms
and of the many parameters  of the semiempirical models could produce model
universes fulfilling all other observational constraints.

In this scenario disks are not primordial to spheroids. They rather develop
only later, being accreted by those spheroids that happen to be in a 
permissive environment.
\par\noindent
{\it Acknowledgements}: 
A.R. would like to thank George Djorgovski, and Ivan King for their 
kind invitation to attend this inspiring and entertaining meeting in honor
of Hy Spinrad. A.R. would also like to acknowledge the support of the Local 
Organization Committee.


\medskip
\noindent
\begin{references}
\reference Aragon-Salamanca, A., Ellis, R.S., Couch, W.J. \&
     Carter, D. 1993, MNRAS, 262, 764 
     Guarnieri, M.D. 1999, A\&A, 341, 539
\reference Bender, R., Saglia, R.P., Ziegler, B., Belloni, P., Greggio, L.,
     Hopp, U. \& Bruzual, G.A. 1997, ApJ, 493, 529
\reference Benitez, N. et al., 1999, ApJL, 515, L65
\reference Bernardi, M., Renzini, A., da Costa, L.N., Wegener, G.,
     et al. 1998, ApJ, 508, L43
\reference Bower, R.G., Lucey, J.R. \& Ellis, R.S. 1992, MNRAS, 254,
             613
\reference Broadhurst, T., \& Bouwens, R.J. 1999, astro-ph/9903009
\reference Broadhurst, T., Ellis, R.S., \& Glazebrook, K. 1992, Nature,
             355, 55
\reference Calzetti D. 1997, in The Ultraviolet Universe at Low and High
Redshift: Probing the Progress of Galaxy Evolution, ed. W.H. Waller,
M.N. Fanelli, J.E. Hollis, \& A.C. Danks, AIP Conference Proceedings
408, (New York: Woodbury), 403
\reference Cimatti A., Andreani P., R\"ottgering H., Tilanus R. 1998,
Nature, 392, 895
\reference Cimatti A., Daddi E., di Serego Alighieri S., Pozzetti L.,
Mannucci F., Renzini A., Oliva E., Zamorani G., Andreani P.,
R\"ottgering H.J.A. 1999, A\&A, submitted
\reference Cohen, J.G. et al. 1999, ApJ, 512, 30
\reference Colless, M., Burstein, D., Davies, R.L., McMahan, R.K., Saglia, 
           R.P., \& Wegner, G. 1999, MNRAS, 305, 259
\reference Cowie, L.L., et al. 1996, AJ, 112, 839
\reference de Carvalho, R. R., \& Djorgovski, S. 1992, ApJ, 389, L49
\reference Dey A., Graham J.R., Ivison R.J., Smail I., Wright G.S. 1999,
ApJ, 519, 610
\reference Dickinson, M. 1995, in Fresh Views of Elliptical Galaxies,
         ed. A. Buzzoni, A. Renzini, \& A. Serrano, ASP Conf. Ser. 86,
         283
\reference Dickinson, M. 1997, in Galaxy Scaling Relations,
         ed. L.N. da Costa \& A. Renzini (Berlin: Springer), p. 215
\reference Dickinson, M. 1998, in The Hubble Deep Field,
ed. M. Livio, S.M. Fall, \& P. Madau (Cambridge: CUP), p. 219
\reference Dunlop, J.S. 1998, astro-ph/9801114
\reference Ellis, R.S., Smail, I., Dressler, A., Couch, W.J., Oemler,
     A. Jr., Butcher, H., \& Sharples, R.M. 1997, ApJ, 483, 582
\reference Elston R., Rieke G.H., Rieke M. 1988, ApJ, 331, L77
\reference Franceschini, A., et al. 1998, ApJ, 506, 600
\reference Graham, J.R., Dey, A. 1996, ApJ, 471, 720
\reference Hu E.M., Ridgway S.E. 1994, AJ, 107, 1303
\reference Jablonka, P., Martin, P., \& Arimoto, N. 1996, AJ, 112, 1415
\reference Jablonka, P., et al. 1999, ApJ, 518, 627
\reference Jimenez, R., Friaca, A.C.S., Dunlop, J.S., Terlevich, R.J.,
           Peacock, J.A., \& Nolan, L.A. 1999, MNRAS, 305, 16
         MNRAS, 280, 167
\reference Jorgensen, I. 1997, MNRAS, 288, 161
\reference Kauffmann, G. 1996, MNRAS, 281, 487
\reference Kauffmann, G., Charlot, S. 1998a, MNRAS, 294, 705
\reference Kauffmann, G., Charlot, S. 1998b, MNRAS, 297, L23
\reference Kauffmann, G., Charlot, S., \& White, S. 1996, MNRAS, 283, 117
\reference Kennicutt R.C. 1998, ARA\&A, 36, 189
\reference Kodama, T., \& Arimoto, N. 1997, A\&A, 320, 41
\reference Kodama, T., Arimoto, N., Barger, A.J., \&
          Aragon-Salamanca, A. 1998, A\&A, 334, 99
\reference Larson, R.B., Tinsley, B.M., \& Caldwell, C.N. 1980, ApJ,
           237, 692
\reference Maraston, C. 1998, MNRAS, 300, 872
\reference Marzke R.O., Da Costa L.N., Pellegrini P.S., Willmer C.N.A., 
Geller M.J.  1998, ApJ, 503, 517
\reference McCarthy P.J., Persson S.E., West S.C. 1992, ApJ, 386, 52
\reference Meneanteau, F. 1998, astro-ph/9811465
\reference Moorwood A.F.M et al. 1999, The Messenger, 95, 1
\reference Newsam A.M., McHardy L.M., Jones L.R., Mason K.O. 1997, MNRAS, 
292, 378
\reference Ortolani, S. Renzini, A., Gilmozzi, R., Marconi, G.,
       Barbuy, B., Bica, E., \& Rich, R.M. 1995, Nature, 377, 701
\reference Pahre, M.A., Djorgovski, S.G., \& de Carvalho, R.R. 1997, in
     Galaxy Scaling Relations: Origins, Evolution and Applications,
     ed. L. da Costa \& A. Renzini (Berlin: Springer), p. 197
\reference Renzini, A. 1994, in Galaxy Formation, ed. J. Silk \& N. Vittorio
           (Amsterdam : North Holland), p. 303
\reference Renzini, A. 1997, ApJ, 488, 35
\reference Renzini, A. 1998b, AJ, 115, 2459
\reference Renzini, A. 1999, astro-ph/9902108
\reference Renzini, A., \& Ciotti, L. 1993, ApJ, 416, L49
             Conf. Ser. 92, 544
\reference Shade, D., Lilly, S.J., Crampton, D., Ellis, R.S., Le F\`evre, O.,
     et al. 1999, astro-ph/9906171
\reference Soifer B.T., Matthews K., Neugebauer G., Armus L., Cohen J.G., 
Persson S.E. 1999, AJ, in press (astro-ph/9906464)
\reference Somerville, R.S., \& Primack, J.R. 1998, astro-ph/9811001
\reference Spinrad, H., Dey, A., Stern, D., Dunlop, J., Peacock, J.,
     Jimenez, R., \& Windhorst, R. 1997, ApJ, 484, 581
\reference Stanford, S.A., Eisenhardt, P.R., \& Dickinson, M. 1998,
    ApJ, 492, 461
\reference Steidel, C. C., Adelberger, K. L., Giavalisco, M.,
     Dickinson, M., \& Pettini, M. 1999, ApJ, 519, 1
\reference Stiavelli, M., Treu, T., Carollo, C.M., Rosati, P.,
           Viezzer, R., Casertano, S., et al. 1999, A\&A, 343, L25
\reference Totani, T., \& Yoshii, Y. 1998, ApJ, 501, L177
\reference Taylor, A.N., Dey, S., Broadhurst, T.J., Benitez, N., \&
     van Kenpen, E. 1998, ApJ, 501, 539
\reference Thompson D., Beckwith S.V.W., Fockenbrock R., Fried J., 
Hippelein H., Huang J.-S., von Kuhlmann, Ch. Leinert, Meisenheimer K., 
Phleps S., R\"oser H.-J., Thommes E., Wolf C. 1999, ApJ, in press
(astro-ph/9907216)
\reference van Dokkum, P. G., Franx, M., Kelson, D. D.,
     \& Illingworth, G. D. 1998, ApJ, 504, L17
\reference Visvanathan, N., \&  Sandage, A. 1977, ApJ, 216, 214
\reference Zepf, S. 1997, Nature, 390, 377
\end{references}
\end{document}